# Could the Vegard strains govern extrinsic size effects in nanoparticles?

Anna N. Morozovska[1*], Iryna S. Golovina[2†], Sergiy V. Lemishko[2,3],

Alexander A. Andriiko[3], Sergiy A. Khainakov[4], and Eugene A. Eliseev[5],

[1] Institute of Physics, National Academy of Sciences of Ukraine,

46, pr. Nauki, 03028 Kyiv, Ukraine

[2] Institute of Semiconductor Physics, National Academy of Sciences of Ukraine,

41, pr. Nauki, 03028 Kyiv, Ukraine

[3] National Technical University of Ukraine ''KPI'', 03056 Kyiv, Ukraine

[4] Department of Organic and Inorganic Chemistry, University of Oviedo, 33006 Oviedo, Spain

[5] Institute of Problems for Material Sciences, National Academy of Sciences of Ukraine,

3, Krjijanovskogo, 03068 Kyiv, Ukraine

**Abstract**

By changing the size and the shape of ferroelectric nanoparticles, one can govern their polar properties including their improvements in comparison with the bulk prototypes. At that the shift of the ferroelectric transition temperature can reach as much as hundreds of Kelvins. Phenomenological description of these effects was proposed in the framework of Landau-Ginsburg-Devonshire (LGD) theory using the conceptions of surface tension and surface bond contraction. However, this description contains a series of poorly defined parameters, which physical nature is ambiguous. It is appeared that the size and shape dependences of the phase transition temperature along with all polar properties are defined by the nature of the size effect. Existing LGD-type models do not take into account that defects concentration strongly increases near the particle surface. In order to develop an adequate phenomenological description of size effects in ferroelectric nanoparticles, one should consider Vegard strains (local lattice deformations) originated from defects accumulation the near surface.

In the paper we propose a theoretical model that takes into account Vegard strains and perform a detailed quantitative comparison of the theoretical results with experimental ones for quasi-spherical nanoparticles, which reveal the essential (about 100 K) increase of the transition temperature in spherical nanoparticles in comparison with bulk crystals. The average radius of nanoparticles was about 25 nm, they consist of $KTa_{1-x}Nb_xO_3$ solid solution, where $KTaO_3$ is a quantum paraelectric, while $KNbO_3$ is a ferroelectric. From the comparison between the theory and experiment we unambiguously established the leading contribution of Vegard strains into the extrinsic size effect in ferroelectric nanoparticles. We determined the dependence of

---

[*] Corresponding author 1: anna.n.morozovska@gmail.com
[†] Corresponding author 2: isgolovina@ukr.net

Vegard strains on the content of Nb and reconstructed the Curie temperature dependence on the content of Nb using this dependence. Appeared that the dependence of the Curie temperature on the Nb content becomes non-monotonic one for the small (< 20 nm) elongated KTa$_{1-x}$Nb$_x$O$_3$ nanoparticles. We established that the accumulation of intrinsic and extrinsic defects near the surface can play the key role in the physical origin of extrinsic size effect in ferroelecric nanoparticles and govern its main features.

**I. Introduction**

The study of unique physical properties of ferroelectric nanoparticles attracts the permanent attention of researchers. Yadlovker and Berger [1, 2, 3] present the unexpected experimental results, which reveal the enhancement of polar properties of cylindrical nanoparticles of Rochelle salt. The authors of the Refs. [4, 5] and [6] perform an excellent ability to manage the temperature of the ferroelectric phase transition, the magnitude and position of the maximum of the dielectric constant for nanopowders and nanoceramics of barium titanate and lead titanate. The studies of KTaO$_3$ nanopowders [7, 8, 9, 10] and nanograined ferroelectrics of KNbO$_3$ and KTa$_{1-x}$Nb$_x$O$_3$ [11, 12, 13] discover the appearance of new polar and magnetic phases, the shift of phase transition temperature in comparison with bulk crystals.

Theoretical consideration of manifold size effects allows one to establish the physical origin of the transition temperature shift and phase diagrams changes appeared under the decrease of nanoparticles sizes. In particular, using the continual phenomenological approach Niepce [14], Huang et al [15], Morozovska et al [16, 17, 18, 19, 20] and Ma [21] have shown, that the changes of the transition temperatures, the enhancement or weakening of polar properties are conditioned by the different "extrinsic" and "intrinsic" size effects in nanoparticles. The partition is tentative and specified by the size effects manifestation. Size effects classification in ferroelectric nanoparticles is given in **Tab. 1**.

As a rule, the term "extrinsic size effect" implies that its consequences depend on the size and the shape of particle, but not on its internal state (e.g. do not depend on the gradients of physical properties inside the nanoparticle). The contribution of the extrinsic size effects leads to the shift of the transition temperature from paraelectric to ferroelectric phase that is proportional to either $1/R$ [14, 16-20] or $1/R^2$ [15] depending on the model, where $R$ is the curvature radius of the nanoparticles surface (e.g. it is the radius of spherical particle). For instance, if one considers intrinsic surface stress (see e.g. [22]) under the curved surface of solid bodies it leads to isotropic compression of the particles resulting in the shift of the transition temperature proportional to $Q\mu/R$, where $\mu$ is the coefficient of the surface stress (similar to the surface tension coefficient determining the surface energy in liquids). The form of the "effective" electrostriction constant $Q$ essentially depends on the shape of the particle, having different signs for the cylinders and spheres of perovskite ferroelectrics [16-20]. In the "surface bound contraction" model [15] the shift of the transition temperature is proportional to the ratio $\chi/R^2$, and the value of the factor $\chi$ is determined by surface bond contraction to the lattice parameter ratio, $\delta a/a$ [**Tab. 1**]. The influence of extrinsic size effects is essential for the particles with the curvature radius (or size) smaller than 50 – 100 nm.

The known "intrinsic size effects" of ferroelectric nanoparticles are determined mainly by the long-range gradient of depolarizing electric field inside the particle and short-range (or "chemical") polarization gradient near the surface [23]. They lead to the more complicated dependence of the transition temperature on the shape and size of the particles, primarily due to a non-trivial dependence of the electrical depolarization fields and flexoelectric strains on particle shape, the orientation of the ferroelectric polarization and the conditions of its screening near the particle surface [16-20]. As a rule, the influence of intrinsic size effects is essential for the particles with the sizes less than 10 nm, the internal scale of polarization is determined by the correlation length, which is typically less than 0.5 nm in spherical particles due to the depolarization effect; the deviation from the bulk polarization is governed by the so called extrapolation length $\lambda$ that is about 0.5 – 2 nm [24, 25].

Table 1. Size effects classification in ferroelectric nanoparticles

| Contribution to size effect (*and its origin*) | Size dependence of Curie temperature shift and dependence on the particle radius $R$ | | Size effect type | Ref. |
|---|---|---|---|---|
| | **Spherical particle of radius $R$** | **Ellipsoid or wire of radius R with polarization directed along the longer axes L>>R** | | |
| **Surface stress** (*originated from the surface tension*) | $-\dfrac{2\mu(2Q_{12}+Q_{11})}{\alpha_T R}$ | $-\dfrac{4\mu Q_{12}}{\alpha_T R}$ | Extrinsic (~ 1/R) | 16, 17 |
| | $\mu$ is a surface stress (tension) coefficient, that is positive, $Q_{ij}$ are electrostriction coefficients, $\alpha_T$ is temperature coefficient of dielectric stiffness | | | |
| **Surface bond contraction** (*originated from the surface curvature*) | $\dfrac{4Y}{\alpha_T}\dfrac{\delta a}{a}\dfrac{(na)^2}{R^2}$ | $\dfrac{2Y}{\alpha_T}\dfrac{\delta a}{a}\dfrac{(na)^2}{R^2}$ | Extrinsic (~ 1/$R^2$) | 15 |
| | $Y$ is a Young modulus, $\delta a$ is contraction of a lattice constant $a$, $n$ is the number of "contracted" layers. Factor $\chi=\dfrac{4Y}{\alpha_T}\dfrac{\delta a}{a}(na)^2$. | | | |

| **Vegard strains** (*originated from defect accumulation*) | $\dfrac{-2\eta(Q_{11}+2Q_{12})}{\alpha_T(s_{11}+2s_{12})}\dfrac{R_0^2}{R^2}$ | $\dfrac{-4\eta Q_{12}}{\alpha_T(s_{11}+s_{12})}\dfrac{R_0^2}{R^2}$ | Extrinsic ($\sim 1/R^2$) | This work |
|---|---|---|---|---|
| | $\eta$ is a Vegard strain, $R_0$ is the defect layer thickness, $s_{ij}$ are elastic compliances | | | |
| **Depolarization, correlation and spontaneous flexoelectric effects** (*originated from electric field and polarization gradients*) | $\dfrac{-(g/\alpha_T)}{(g/n_d)+(\lambda+\sqrt{g/n_d})(R/3)}$ $\lambda > 0$ $n_d = \dfrac{1}{(\varepsilon_b + 2\varepsilon_e)\varepsilon_0}$ is a depolarization factor | $-\dfrac{2}{\alpha_T}\left(\dfrac{g}{R\lambda + R^2/4}\right),\ \lambda > 0,$ $-\dfrac{2}{\alpha_T}\left(g\dfrac{2\lambda - R}{2R\lambda^2}\right),\ \lambda < 0.$ The shift from depolarization field is negligibly small, absent in the limit of a wire when $L/R \to \infty$ | Intrinsic (R-dependence is complex) | 16, 17, 26 |
| | $\lambda$ is an extrapolation length, g is the polarization gradient term, $n_d$ is a depolarization factor, $\varepsilon_b$ is a background permittivity of ferroelectric [27], $\varepsilon_e$ is the permittivity of external media. | | | |

Therefore the analysis of the experimentally observed transition temperature dependence on the particles sizes allows one to establish the nature of size effects in the studied system and to determine corresponding phenomenological parameters like $\mu$, $\chi$ or $\lambda$ from the fitting of experimental data with the adequate theoretical model. Despite the goal seems clear, it is still to be realized for the majority of nanostructured systems. Probably, the difficulties are mostly due to the fact that existing phenomenological considerations of ferroelectric particles are not adequate for real nanoparticles with strongly strained near-surface layers, where the strains or stresses are caused by the accumulation of defects (impurities and vacancies) in the region. Indeed, it regarded well-established that the defects concentration noticeably increases near the particle surface allowing for the essential lowering of their formation energies [28, 29, 30]. The spontaneous polarization abrupt near the surface of ferroelectric causes the strong accumulation of ions and charged vacancies in the spatial regions to screen bond surface charges [31, 32]. In turn, vacancies and ions accumulation near the surfaces of solids produces strains that substantially alter thermodynamic equilibrium [33, 34], which in turn lead to the changes of phase diagrams and transition temperatures.

In order to develop an adequate thermodynamic description of size effects in ferroelectric nanoparticles one has to determine a microscopic nature of phenomenological parameters and relate them with the lattice deformation near the particle surface due to the defects accumulation. Thorough analysis of our experimental results has shown that the conception of "chemical" pressure [35] originated from the Vegard strains [36, 37] perfectly suits to our aim. According to this conception, the inclusion of a defect (impurity ion or vacancy) leads to the local deformation of the crystal lattice, and the action of many defects causes the strain proportional to their concentration. In the case, the proportionality coefficient is determined either from the *ab-initio* calculation [37] or from experiments [38, 39]. It should be noted that the influence of Vegard strain, coming from the diffusion and the accumulation of defects near the interfaces of ferroelectric thin films, results in the pronounced change of their polar properties [40, 41]. Therefore it is natural to expect that one could not neglect Vegard strains when describing polar properties of ferroelectric nanoparticles. At that a steric effect [42, 43] takes place for strong accumulation ("crowding") of defects.

In order to realize this conception we performed the modeling of the transition temperatures for ferroelectric nanoparticles with taking into account the Vegard strains. Theoretical results are analyzed and compared with experiment for $KTa_{1-x}Nb_xO_3$ nanoparticles.

## II. Theoretical model
### II.1. Model background for $KTaO_3$-$KNbO_3$ system

It was established experimentally that extrinsic and intrinsic defects play a crucial role in the emerging of polar properties in nanocrystals of $KTa_{1-x}Nb_xO_3$ [9, 12-14]. Thus, due to the presence of the dipole centers accosiated with the $Fe^{3+}$ ions (the centers of axial and rhombic symmetries) ferroelectric phase transition occurs in nanocrystalline $KTaO_3$ at 29 K [7, 8], while the bulk material is a quantum paraelectric [44]. These centers include oxygen vacancies in their structure and are formed mainly near the surface of the particles, as the concentration of oxygen vacancies have a sharp maximum in the surface layer.

Let us remind the main features of polar and structural phase transitions in the considered system. Bulk $KTaO_3$ is a quantum paraelectric having cubic structure down to 0 K [44]. At the same time $KNbO_3$ and solid solutions of $KTa_{1-x}Nb_xO_3$ ($x>0.2$) undergo three successive phase transitions, namely from cubic paraelectric to tetragonal ferroelectric phase at Curie temperature ($T_C$) and then two structural transitions with the polarization vector *P* switching between different crystallographic directions [**Fig. 1a**]. At that, as it was shown in Ref. [45], $T_C$ shifts almost linearly under the changes of Nb content *x* in $KTa_{1-x}Nb_xO_3$ solid solution. Below we aim to consider the dependence of the Curie temperature on the composition under decreasing the size of the crystals for two particle shapes.

In the considered solid solutions the most prevalent intrinsic defects are vacancies of oxygen and potassium [**Fig. 1b**]. The ions of iron and manganese are mostly registered among the extrinsic defects [**Figs 1c** and **1d**], which are incorporated into the lattice during the synthesis as unavoidable impurities [9, 13, 49]. It is known that the concentration of defects inside the crystal is inhomogeneous and strongly increases when going from the bulk to the surface of the particle [49].

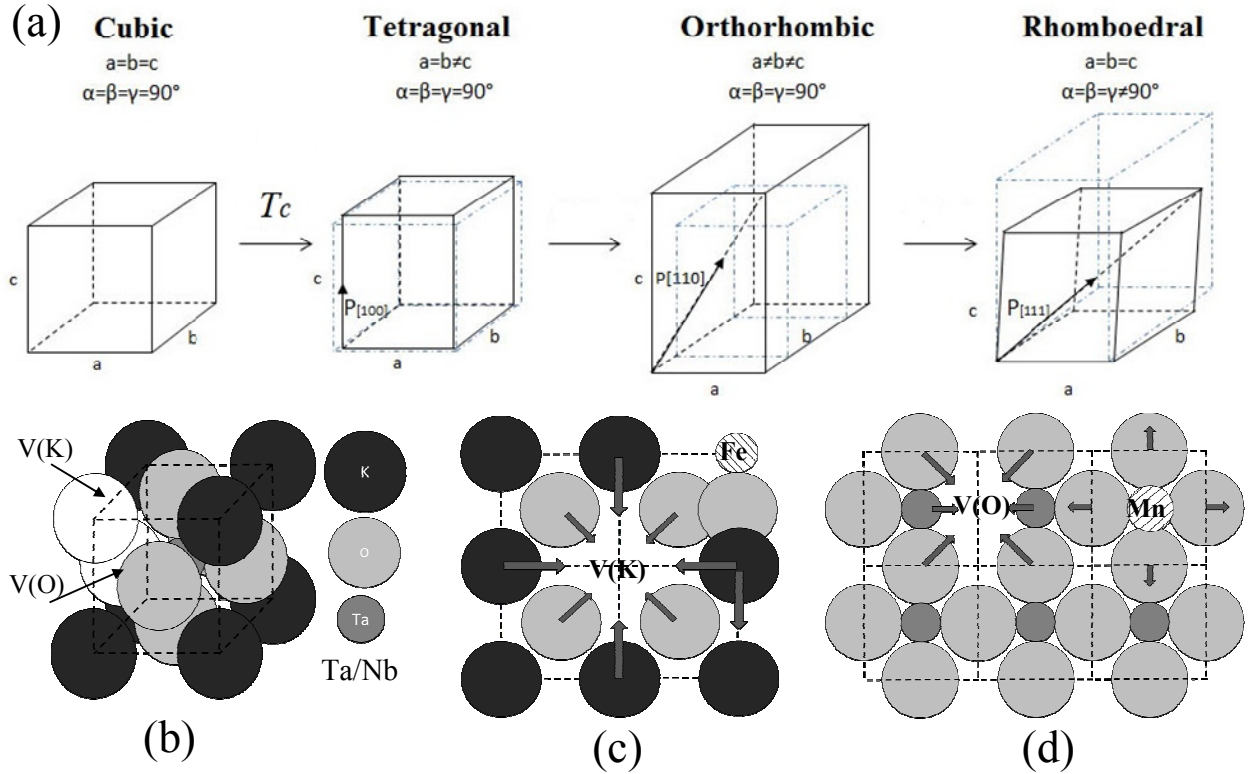

**Figure 1. (a)** Sequence of phase and structural transformations in $KTa_{1-x}Nb_xO_3$. **(b)** Intrinsic defects in KTa(Nb)O$_3$ lattice. **(c, d)** Local lattice deformations caused by intrinsic (oxygen and potassium vacancies - V(O) and V(K)) and extrinsic ($Fe^{3+}$ or $Mn^{2+}$ ions) defects.

Since the near-surface layer is enriched with defects, here the crystal lattice becomes either "spongy" or "denser" depending on the type of defect [**Figs 1b-1d**], meaning that the lattice parameter is also locally changed. Since the presence of different types of defects, resulting into either expansion or compression of the lattice, the surface of the particle may have a complicated relief. However the assemble of weakly interacting or non-interacting particles most probably can be considered as "effective" assemble of ellipsoids with different aspect ratio of the semi-axes $R$ and $L$. In our calculations we change the curvature of the surface $1/R$ and calculate the Curie temperature for two limiting forms of the particle, namely the sphere and very prolate ellipsoid. So that $R$ is either the sphere radius or the smaller semi-axis of the ellipsoid hereinafter.

**II.2. Basic equations**

Possible lattice deformations caused by intrinsic (vacancies V(O) and V(K)) and extrinsic ($Fe^{3+}$ or $Mn^{2+}$ ions) defects are shown in **Figs 1b-d.** Following Huang et al. [15] and using the conception of Vegard stress [35-37], the defects accumulation under a curved surface produces an effective hydrostatic pressure of the inner part of the particle. The screening of depolarization field at the surface and outside of the particle leads to the exponential decrease of charged defects concentration when moving away from the surface. At that, the characteristic thickness of the layer enriched by defects is determined by the screening length, and their maximal concentration is limited by steric effect [40, 41]. Our calculations have shown that the existence of well-localized defects layer leads to the hydrostatic pressure acting on the inner part of ellipsoidal particle with semi-axes $R$ and $L$:

$$\sigma_{rr}(R,x) = \begin{cases} \dfrac{-\eta(x)}{s_{11}(x)+s_{12}(x)} \dfrac{R_0^2(x)}{R^2}, & R \ll L \quad (prolate \;\; ellipsoid), \\ \dfrac{-\eta(x)}{s_{11}(x)+2s_{12}(x)} \dfrac{R_0^2(x)}{R^2}, & R \approx L \quad (sphere). \end{cases} \qquad (1)$$

Here subscript "$rr$" denotes that the pressure is radial, $R$ is the particle radius. $s_{ij}(x)$ are elastic compliances modulus of the material, the characteristic size $R_0(x)$ is the particle surface layer thickness, where accumulated defects create elementary volume changes. Parameter $\eta(x)$ is a "chemical" Vegard strain that is usually dependent on the Nb content $x$ in $KTa_{1-x}Nb_xO_3$ due to the strong dependence of the elementary volume changes on defect and surrounding atoms type [37]. Immediately the Vegard strain $x$-dependence can lead to the analogous dependence of the shell thickness, $R_0(x)$. Moreover the following correlation should exist: the higher is the Vegard strain, the thicker should be the shell, because more numbers of layers are required for the full relaxation of the surface strain. Let us underline that the stress $\sigma_{rr}$ in Eq.(1) is radius and content dependent; but its Nb content dependence is preliminary unknown for $KTa_{1-x}Nb_xO_3$ and will be determined from our experimental data. Typical value of $\sigma_{rr}$ is about $10 \left(R_0^2/R^2\right)$GPa for a Vegard strain of about 1% and elastic compliances of about $10^{-12}$ Pa, so this is enough high value. With $R$ increase the stress decreases proportional to $\left(R_0^2/R^2\right)$.

Since the actual particle sizes for which we have experimental data are higher than 10 nm [**Fig. 2e**], below we can ignore intrinsic size effects, so LGD-potential functional for the solid solution of paraelectric and ferroelectric acquires relatively simple form [16 - 20]:

$$F = \alpha(T,R,x)\frac{P^2}{2} + \beta(x)\frac{P^4}{4} + \gamma(x)\frac{P^6}{6} - PE \qquad (2)$$

Here $x$ is the content of Nb in $KTa_{1-x}Nb_xO_3$. $P$ stands for polarization, $E$ is electric field. Coefficient $\alpha(T,R,x)$ depends on temperature $T$, particle size $R$, content $x$, polarization orientation and other material parameters.

Using direct variational method [16 - 20] for ellipsoidal particles with semi-axes $R$ and $L$, which polarization is uniformly aligned along the longer ellipsoid axis $L$, the coefficient $\alpha(T,R,x)$ becomes renormalized by the stress $\sigma_{rr}(R,x)$ given by Eq.(1) via electrostriction effect, $\alpha(T,R,x) = \alpha_{bulk}(T,x) + Q_{ij}(x)\sigma_{ij}(R,x)$, where $Q_{ij}$ are electrostriction tensor coefficients [16-17]. So that the coefficient was calculated as:

$$\alpha(T,R,x) = \begin{cases} x\alpha_{Tf}(T-T_C^b) + (1-x)\alpha_{Tq}\left(\frac{T_q}{2}\coth\left(\frac{T_q}{2T}\right) - T_0\right) + \frac{4\eta(x)Q_{12}(x)}{s_{11}(x) + s_{12}(x)}\frac{R_0^2}{R^2}, & R \ll L, \\ x\alpha_{Tf}(T-T_C^b) + (1-x)\alpha_{Tq}\left(\frac{T_q}{2}\coth\left(\frac{T_q}{2T}\right) - T_0\right) + \frac{2\eta(x)(Q_{11}(x) + 2Q_{12}(x))}{s_{11}(x) + 2s_{12}(x)}\frac{R_0^2}{R^2}, & R \approx L. \end{cases} \qquad (3)$$

Hereinafter subscripts "$q$" and "$f$" denotes the values related with proper ferroelectrics and quantum paraelectrics, respectively. We used that $KTa_{1-x}Nb_xO_3$ parent phase has cubic m3m symmetry. Here $T_C^b$ is the ferroelectric Curie temperature of bulk material. In Eqs.(3) we used the Barrett-type formula for the coefficient $\alpha_q(T,R) = \alpha_{Tq}((T_q/2)\coth(T_q/2T) - T_0)$, which is valid for quantum paraelectrics [46] in a wide temperature interval including low quantum temperatures. $T_0$ and $T_q$ are extrapolated "virtual" Curie temperature and characteristic quantum oscillations temperatures, correspondingly. At temperatures $T \gg T_q/2$ the Barrett formulae transforms into the classical limit, $\alpha_q(T,R) \approx \alpha_{Tq}(T - T_0)$. Electrostriction tensor coefficients $Q_{ij}$ content $x$ dependence can be regarded linear $Q_{ij}(x) = xQ_{ij}^f + (1-x)Q_{ij}^q$. Linear dependencies can also be used for elastic compliances, $s_{ij}(x) = xs_{ij}^f + (1-x)s_{ij}^q$, Vegard strain $\eta(x) = x\eta_f + (1-x)\eta_q$ and shell thickness, $R_0(x) = xR_f + (1-x)R_q$.

The Curie temperature $T_C(x,R)$ of the solid solution can be determined from the condition $\alpha(T_C,R,x) = 0$ that in the classical limit, $T_C \gg T_q/2$, gives evident analytical expressions

$$T_C(R,x) = \begin{cases} \dfrac{x\alpha_{Tf}T_C^b + (1-x)\alpha_{Tq}T_0}{x\alpha_{Tf} + (1-x)\alpha_{Tq}} - \dfrac{4\eta(x)Q_{12}(x)(R_0^2/R^2)}{(s_{11}(x)+s_{12}(x))(x\alpha_{Tf}+(1-x)\alpha_{Tq})}, & R \ll L, \\ \dfrac{x\alpha_{Tf}T_C^b + (1-x)\alpha_{Tq}T_0}{x\alpha_{Tf} + (1-x)\alpha_{Tq}} - \dfrac{2\eta(x)(Q_{11}(x)+2Q_{12}(x))(R_0^2/R^2)}{(s_{11}(x)+2s_{12}(x))(x\alpha_{Tf}+(1-x)\alpha_{Tq})}, & R \approx L. \end{cases} \quad (4)$$

For the case of the second order phase transitions, $T_C(x,R)$ is the transition temperature from ferroelectric to paraelectric phase. For the first order phase transitions the condition $\alpha(T_C, R, x) = 0$ still gives the Curie temperature, but the ferroelectric-paraelectric transition temperature, $T_{FE}(x,R)$, should be found as a solution of transcendental equation $\alpha(T_{FE}, R, x) = 3\beta(x)^2/(16\gamma(x))$. The spontaneous polarization at $E=0$ is $P_0^2 = \left(\sqrt{\beta^2 - 4\alpha\gamma} - \beta\right)/2\gamma$.

As one can see from Eq.(4) the shift of Curie temperature is induced by joint action of Vegard strain and size effects, that is reflected by the product $\eta(R_0/R)^2$. The radius dependence of Curie temperature shift is governed by the ratio $(R_0/R)^2$. Equations (4) show that the increase of Curie temperature in comparison with a bulk material appears under the condition $\eta(x)(Q_{11}(x)+2Q_{12}(x)) < 0$ for spherical particles or $\eta(x)Q_{12}(x) < 0$ for ellipsoidal particles with high aspect ratio $L/R \gg 1$.

### III. Nb-content and size dependence of the nanoparticles transition temperature

To fit the experimental dependence of the Curie temperature $T_C(x)$ (squires in **Fig. 2a**) and ferroelectric transition temperature $T_{FE}(x)$ (triangles in **Fig. 2a**) on the content $x$ of Nb we use the interpolate function for $T_0(x) = T_0(1 + 25(1-x)^{0.7})$. Our fitting based on Eqs.(1)-(4) is shown by solid (for $T_{FE}(x)$) and dotted (for $T_C(x)$) curves. Corresponding parameters of the bulk KTaO$_3$ and KNbO$_3$ are listed in **Tab. 2.**

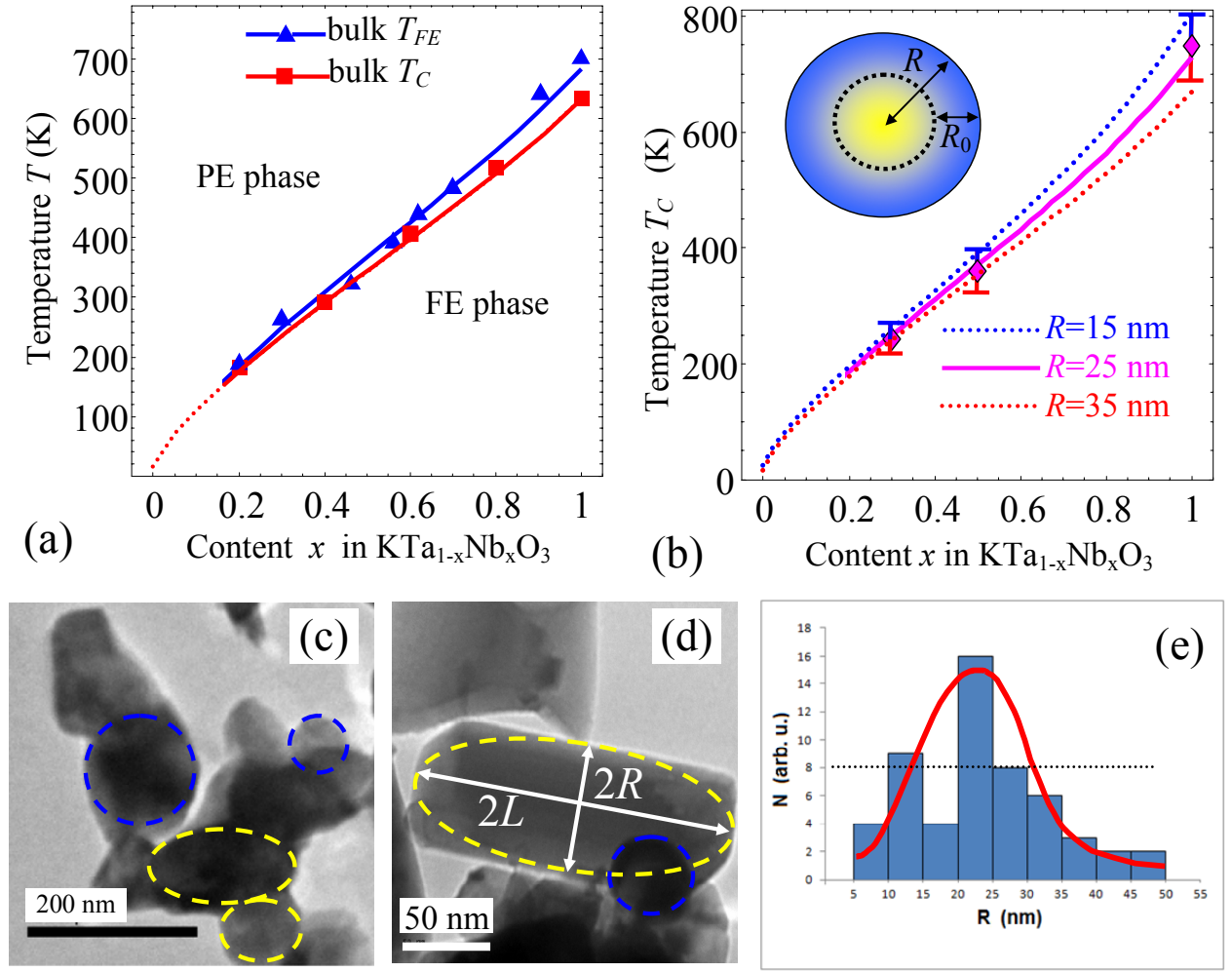

**Figure 2. (a)** Curie and ferroelectric transition temperature vs. Nb content $x$ in bulk $KTa_{1-x}Nb_xO_3$. Symbols are experimental data from Ref.[10], solid curves are theoretical fitting based on Eqs.(5)-(6). Abbreviations PE and FE stand for paraelectric and ferroelectric phases. Parameters of the bulk $KTaO_3$ and $KNbO_3$ are listed in **Tab. 2**. **(b)** Curie temperature vs. Nb content $x$ in assembles of $KTa_{1-x}Nb_xO_3$ nanoparticles. Different symbols are experimental data, solid and dotted curves are theoretical fitting for the nanoparticle with radius 25, 15 and 35 nm, respectively. Fitting parameters are listed in **Tab. 3**. **(c,d)** TEM images which show the quasi-spherical and ellipsoidal $KTa_{1-x}Nb_xO_3$ nanoparticles. **(e)** The size distribution of nanoparticles. Solid red curve is the histogram fitting by polynomial-gaussian fit.

**Table 2.** Parameters of the bulk $KTaO_3$ and $KNbO_3$ extracted from Refs.[10, 45, 47, 48]

| Physical quantity | Quantum paraelectric $KTaO_3$ | Ferroelectric $KNbO_3$ |
|---|---|---|
| Coefficient $\alpha_T$ | $\alpha_{Tq} = 1.36 \times 10^6$ m/(FK) | $\alpha_{Tf} = 4.6 \times 10^5$ m/(FK) |

| Characteristic temperatures | $T_q = 55$ K, $T_0 = 15$ K | $T_C = 633$ K, $T_{FE} = 698$ K |
|---|---|---|
| Electrostriction coefficients | $Q_{11} = 0.11\,\text{m}^4/\text{C}^2$, $Q_{12} = -0.023\,\text{m}^4/\text{C}^2$ | $Q_{11} = 0.13\,\text{m}^4/\text{C}^2$, $Q_{12} = -0.047\,\text{m}^4/\text{C}^2$ |
| Elastic compliances | $s_{11} = 2.7\times 10^{-12}$ Pa$^{-1}$, $s_{12} = -6.25\times 10^{-13}$ Pa$^{-1}$ | $s_{11} = 4.61\times 10^{-12}$ Pa$^{-1}$, $s_{12} = -1.11\times 10^{-12}$ Pa$^{-1}$ |

**Table 2** gives us the parameters of the bulk materials, but the composition dependences of Vegard strain coefficient and shell thickness, which determine the strength of the size effects in nanoparticles in accordance with e.g. Eq.(4) remain unknown.

In order to obtain the dependences we performed the comparison with our experiment. KTa$_{1-x}$Nb$_x$O$_3$ nanoparticles fabrication process is described in details in Refs. [8, 11, 12] and [49]. Examples of TEM images, which show the nanoparticles of quasi-spherical and ellipsoidal shape, one can see in **Figs 2c** and **2d**. Most of the fabricated nanoparticles have quasi-spherical shape; their radii are distributed around the *most probable size of* 20-25 nm (see histogram in **Fig. 2e**). The full width at half maximum (FWHM) of the particle sizes distribution is about 20 nm. Despite there were some sizable amount of the big nanoparticles, they contribute almost nothing to the studied size effects, acting as "bilk" background and thus slightly elongate the error bars of the Curie temperature to lower temperatures. Experimentally, the Curie temperature was determined from Raman scattering measurements using Jobin-Yvon/Horiba T64000 Raman triple spectrometer. The temperature dependences of intensity, width and frequency of relevant modes of the Raman spectra were thoroughly analyzed. In details, the experimental technique was described elsewhere [11, 12].

Comparison of calculated Curie temperature with experimental results for KTa$_{1-x}$Nb$_x$O$_3$ nanoparticles is shown in **Fig. 2b.** We attributed the diamonds in the figure to experimental data for nanoparticles with the most probable radius 25 nm. Error bars show the Curie temperature scattering originated from the particle radii deviation from 25 nm on about ±10 nm, that corresponds to the radii from 15 nm to 35 nm. Solid and dotted curves are theoretical fitting for the 25, 15 and 35 nm, correspondingly. Let us underline the pronounced increase of the Curie temperature for KTa$_{1-x}$Nb$_x$O$_3$ nanoparticles in comparison with a bulk solid solution for $x>0.8$ leading to a strong enhancement of ferroelectric properties. Since $Q_{12}(x)<0$ and $Q_{11}(x)+2Q_{12}(x)>0$ for all $x$, negative Vegard strain $\eta(x)$ increases the Curie temperature for spherical nanoparticles, positive strains $\eta(x)$ increase the temperature for the prolate ellipsoids accordingly to Eqs.(4). Content dependent parameters determined from the fitting to experiment are listed in **Tab. 3.**

**Table 3.** Parameters of the nanosized $KTa_{1-x}Nb_xO_3$

| Experimental data for Curie temperature in $KTa_{1-x}Nb_xO_3$ | | Content dependent parameters determined from the fitting to experimental shown in the Figure 2b | | |
|---|---|---|---|---|
| Content x | Curie temperature | Spheres fraction $g(x)=xg_f+(1-x)g_q$ $g_f=0.95$, $g_q=0.9$ | Shell thickness $R_0(x)=xR_f+(1-x)R_q$ $R_f=4.2$ nm, $R_q=2.4$ nm | Vegard coefficient $\eta(x)=x\eta_f+(1-x)\eta_q$ $\eta_f=-3\%$, $\eta_q=-1\%$ |
| **x=0.3** | $T_C = (243\pm30)$K  bulk $T_C = 238$ K | 0.915 | 2.8 nm | $-1.6\%$ |
| **x=0.5** | $T_C = (359\pm40)$K  bulk $T_C = 351$ K | 0.925 | 3.3 nm | $-2\%$ |
| **x=1** | $T_C = (748\pm50)$K  bulk $T_C = 633$ K | 0.95 | 4.2 nm | $-3\%$ |

We found out that the best fitting of Curie temperature for the quasi-spherical particles of radius $(25 \pm 10)$ nm corresponds to the negative Vegard strain $\eta(x) = -0.01x - 0.03(1-x)$ and shell thickness $R_0(x) = (2.4x + 4.2(1-x))$ nm. The latter values are quite reasonable and correspond to the 6 – 10 unit cell thick shell. As anticipated, effective spheres fraction in the composite appeared rather high, $g(x) = 0.95x + 0.9(1-x)$, but some small amount of prolate ellipsoidal particles (from 5% to 10% depending on the content $x$) exists due to the presence of elongated particles with high aspect ratio $L/R \gg 1$.

Here an important remark should be made. We failed to fit the experimental data using both effective surface tension and intrinsic size effects models [16-17] for the realistic values of the surface stress coefficient, since the fitting required very high negative values of the coefficient (about $-(10\text{-}20)$N/m), that is in contradiction with surface equilibrium and realistic range of the coefficient $+(1\text{-}2)$N/m in different perovskites and other oxides [21]. Inclusion of the intrinsic size effect (polarization gradient along with induced depolarization field) leads to the correct trend that agree with experiment, namely to the Curie temperature increase with radius decrease, only at negative extrapolation lengths and the increase becomes noticeable at the particles radius less than 5 nm. Since there is no solid background for the existence of negative extrapolation length and the minimal particle radius represented in the histogram is 5 nm [**Fig.2e**], including of the intrinsic size effect does not help us to describe the experiment. In contrast, the model based on the Vegard strain appeared in a quantitative agreement with the experimental results for realistic values of all fitting parameters. The surface bond contraction model [15] gives the same radius dependence of the Curie temperature shift than the Vegard strain-based model, but it does not allow quantitative fitting the experimental data dependence on Nb content $x$, since only the Young modulus $Y$ is $x$-dependent in the factor $\chi = \dfrac{4Y}{\alpha_T}\dfrac{\delta a}{a}(na)^2$ included in the model (see **Tab. 1**). Substitution of the known dependence Y(x) does not lead to a reasonable fitting. Other factors like the surface bond contraction and the lattice parameter are almost content-independent. Probably the surface bond contraction model is adequate for the nanoparticles without surface defects.

Since we extract the dependence of the Vegard strain and shell thickness on Nb content $x$ from experimental data in an unambiguous way, we can enough reliably reconstruct the impact of size effects on the nanosized $KTa_{1-x}Nb_xO_3$ using these dependencies and vary the particle radius and the sphere-to-ellipsoid ratio in the particles assemble, because these factors can be controlled in realistic experiments. Such procedure opens the opportunity to predict the Vegard strain impact on the extrinsic size effects in nanoparticles. So, let us consider a $KTa_{1-x}Nb_xO_3$ nanocomposite, where particles sizes are distributed around the average value, and the particle shape varies, namely there are some fraction of ellipsoidal and spherical particles in the material, but electric and elastic interaction between the particles can be regarded small due to the screening effects and surface stresses in the interfacial regions. Using Eqs.(1)-(4) and parameters listed in **Tabs 2-3**, we study content and radius dependences of the extrinsic size effect in the composites with the fraction of spheres varying from 100% to 0%. Reconstructed dependencies of the Curie temperature on Nb content $x$ and particle radius $R$ are depicted in **Figs 3** and **4**.

Curie temperature increases with the increase of Nb content in bulk $KTa_{1-x}Nb_xO_3$ (dotted curves in **Figs 3**). Unexpectedly the Curie temperature $T_C(x,R)$ non-monotonically depends on Nb content $x$ in assembles of small ($R$=15 nm) $KTa_{1-x}Nb_xO_3$ nanoparticles, when the fraction of spheres becomes less than 50 % for negative Vegard strain $\eta(x) = -0.01x - 0.03(1-x)$ [**Fig. 3a**]. Such unexpected behavior could not be described as the consequence of particle shape changes, since the latter was constant under the increase of $x$. At the same time the change of composition leads to the changes of the terms $\dfrac{4\eta(x)Q_{12}(x)}{s_{11}(x)+s_{12}(x)}\dfrac{R_0^2(x)}{R^2}$ and $\dfrac{2\eta(x)(Q_{11}(x)+2Q_{12}(x))}{s_{11}(x)+2s_{12}(x)}\dfrac{R_0^2(x)}{R^2}$ in Eqs.(3)-(4), proportional to the Vegard strain $\eta(x)$, which changes in a most strong way with composition $x$ according to **Tab. 3**. Thus the non-monotonic behavior is a direct sequence of the rather strong increase of the Vegard strain with $x$ increase. For the case when the fraction of spheres is higher than 50%, Curie temperature monotonically and super-linearly increases with $x$ increase and can overcome the bulk Curie temperature on hundreds of Kelvins for $x>0.8$.

**Figure 3b** illustrates how the Curie temperature changes for positive Vegard strain, $\eta(x) = 0.01x + 0.03(1-x)$, and the same other parameters as in **Fig. 3a**. Qualitatively, the behavior for positive $\eta(x)$ is complementary to the one for negative $\eta(x)$ (compare **Figs 3a** and **3b**). Namely, for $\eta(x) > 0$ sub-linear $x$-dependence of $T_C$ appears when the fraction of prolate ellipsoids becomes less than 50 %. Curie temperature monotonically and super-linearly increases with Nb content increase and can overcome the bulk Curie temperature for the case when the fraction of spheres is lower than 50 %.

These results approve the statement that the principal behavior of Curie temperature is governed by the Vegard strain sign (compression or tension), its absolute value and particle shape (prolate or spherical one). In particular, one can see from **Figs 3a** and **3b**, which correspond to the opposite signs of the Vegard strain, that for positive $\eta$ values the curves for prolate ellipsoidal nanoparticles are located above the dotted curve corresponding to the bulk material; curves for spherical nanoparticles are located below the dotted curve. The curves sequence is opposite for negative $\eta$ values. The curves order and slope changes with respect to the bulk dotted curve with changing the fraction of spheres. In numbers, the transition temperature in nanosized composite could be from tens to hundred Kelvins lower or higher than in the bulk $KTa_{1-x}Nb_xO_3$.

The smaller is the particle radius $R$, the stronger is the deviation of the curves from the dotted bulk ones [**Fig. 3c**]. Size effects are visible for the radii less than 50 nm, at that the minimal content of Nb required for the ferroelectricity appearance is different for a bulk material (20%) and nanoparticles (from 10 to 30%) depending on the particle size, shape and Vegard strain sign. In the case of negative Vegard strain ($\eta(x) < 0$) the ellipsoids are characterized by non-monotonic dependence of Curie temperature on the Nb content $x$, having a maximum at $x=0.6$ and the second critical concentration of Nb, $x=0.95$, at which ferroelectric order disappears. **Figure 3d** illustrates model situation, when $\eta(x) > 0$ and other parameters are the same as for **Fig. 3c**. One could see from this figure that it is possible to have a non-monotonic dependence of Curie temperature on Nb content for spherical nanoparticles of small radius; for the prolate ellipsoids the Curie temperature monotonically increases with the increase of Nb content with super-linear trend.

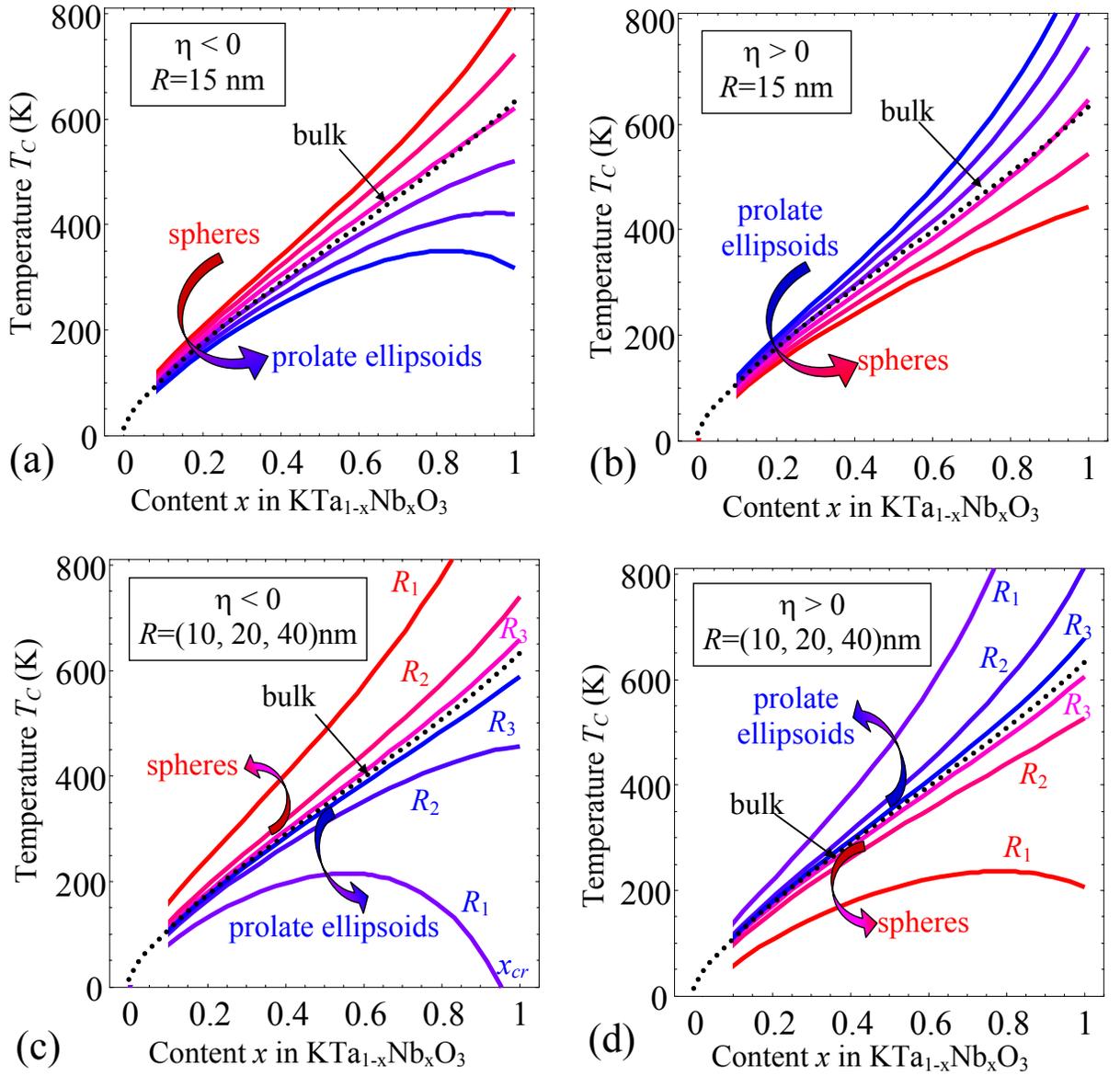

**Figure 3. Curie temperature vs. Nb content $x$ in nanosized $KTa_{1-x}Nb_xO_3$**. Plots are calculated for negative **(a,c)** and positive **(b,d)** Vegard strains $\eta(x) = \pm(0.01x + 0.03(1-x))$. The particle radius is $R = 15$ nm in the plots **(a)** and **(b)**. The spheres fraction changes from 100% to 0% with a step of 20% for different curves in the direction indicated by arrow. Plots **(c)** and **(d)** are calculated for different particle shape (prolate ellipsoids and spheres as indicated by arrows) and radii $R_1 = 10$ nm, $R_2 = 20$ nm, $R_3 = 40$ nm (indicated by labels "$R_i$" near the curves). Material parameters are listed in **Tabs 2-3**.

Reconstructed dependencies of Curie temperature via particle radius are shown in **Figs 4** for different $x$ and spheres-to-ellipsoids ratio. At fixed content $x$ the Curie temperature $T_C$ tends to the bulk value with the particle radius increase. With the radius decrease we see either the strong increase of $T_C$ proportional to $R_0^2/R^2$ for negative product $\eta(2Q_{12}+Q_{11})$ or $\eta Q_{12}$ for spherical or prolate nanoparticles, respectively (see Eqs.(4)), or its rapid decrease up to $T=0$ K at some "critical" radius $R_{cr}$ in the opposite case (compare top and bottom curves in **Figs. 4**). The critical radius $R_{cr}$ of the ferroelectricity disappearance exists for the positive product $\eta(2Q_{12}+Q_{11})$ or $\eta Q_{12}$ for spherical or prolate nanoparticles, respectively. $R_{cr}$ increases with the increase of $\eta$ absolute value.

The strong changes in radius dependence of the Curie temperature $T_C$ appeared when the fraction of spheres varies from 100% to 0%. Actually, in the composite with compressed spherical nanoparticles ($\eta<0$) $T_C$ increases with the radius decrease. With increasing the fraction of prolate ellipsoids $T_C$ gradually decreases, then becomes lower than the bulk one and rapidly decreases with the radius decrease in the composite with 50, 20 and 0% of spheres. Note, that the situation is visa versa for particles under tensile strains, $\eta>0$; here the temperature increases with the fraction of prolate particles increase.

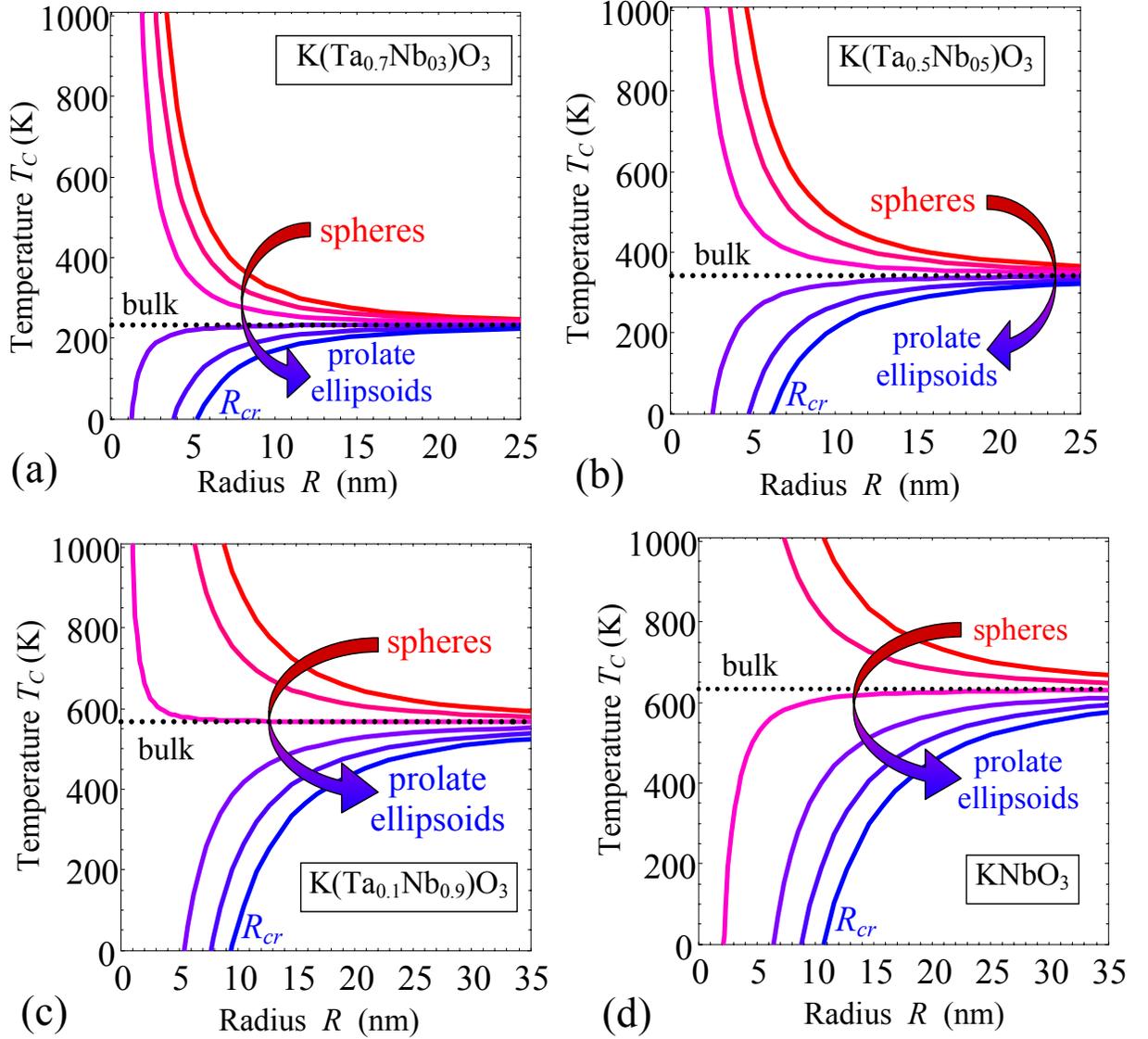

**Figure 4.** Curie temperature vs. particle radii $R$, calculated for Vegard strain $\eta(x) = -0.01x - 0.03(1-x)$ at Nb content $x=0.3$ **(a)**, $x=0.5$ **(b)**, $x=0.9$ **(c)**, and $x=1$ **(d)**. The spheres fraction changes from 100% to 0% with a step of 20% for different curves in the direction indicated by arrow. Material parameters are listed in **Tabs 2 - 3**.

### IV. Summary

We proposed a phenomenological description of size effects in ferroelectric nanoparticles taking into account Vegard strains caused by defects accumulation near the surface of the particle. Performed calculations and detailed quantitative comparison with experimental results on quasi-spherical nanoparticles of $KTa_{1-x}Nb_xO_3$ solid solution allow determining unambiguously the key impact of Vegard strain on the extrinsic size effect and revealing the essential (about 100 K) increase of the transition temperature in spherical nanoparticles in comparison with bulk crystals.

We also determined the dependence of Vegard strains on the Nb content and using this dependence reconstructed the content dependence of the solid solution Curie temperature. Appeared that the dependence of the Curie temperature on the Nb content should be non-monotonic for the small elongated $KTa_{1-x}Nb_xO_3$ nanoparticles (at size < 30 nm).

In this way it is established that the key role in the origin of extrinsic size effects in nanoparticles belongs to the accumulation of intrinsic and extrinsic defects near the surface that cause local Vegard strains. The strains can govern the main features of the particles ferroelectric properties.